# Towards on-chip nascent all-van-der-Waals polarization optical components for nanoscale photonic applications

Dmitrii Litvinov[1, 2, †], Ilia Begichev[1, 2, †], Iakov Reznikov[1, 2], Changyi Chen[1, 2], Maksim Sargsyan[3], Sergey Y. Grebenchuk[1], Makars Siskins[1, 4], Kostya S. Novoselov[1, 2], Alexey I. Berdyugin[1, 2, 5], Maciej Koperski[1, 2, *] & Davit A. Ghazaryan[1, 3, *]

*[1]Institute for Functional Intelligent Materials, National University of Singapore, Singapore 117544, Republic of Singapore*
*[2]Department of Materials Science and Engineering, National University of Singapore, Singapore 117575, Republic of Singapore*
*[3]Laboratory of Advanced Functional Materials, Yerevan State University, Yerevan 0025, Republic of Armenia*
*[4]School of Physics and Astronomy, University of Southampton, Highfield, Southampton SO17 1BJ, United Kingdom*
*[5]Department of Physics, National University of Singapore, Singapore 117575, Republic of Singapore*

*†These authors contributed equally to this work*

**The correspondence should be addressed to: msemaci@nus.edu.sg & davitghazaryan@ysu.am*

**The integration of polarization-control elements into nanoscale photonic circuits remains a central challenge for modern on-chip optical technologies, which call for continuous miniaturization. Van der Waals (vdW) crystals provide a versatile platform for the creation of such components owing to their strong optical anisotropy, high-refractive indices and atomically precise heterostructure assembly capabilities. In this work, we demonstrate an approach towards the creation of all-vdW on-chip polarization optical components, exemplified with quarter-wave plates operating in the near-infrared (NIR) spectral region, realized in specifically twisted $ReSe_2/\alpha\text{-}MoO_3$ vdW heterostructures on a $Si/SiO_2$ platform. Here, low-symmetry $ReSe_2$ layer serves as a source of exceptionally high linearly polarized excitonic emission, whereas in-plane birefringent $\alpha\text{-}MoO_3$ provides polarization-state conversion. Furthermore, our finite-difference time-domain (FDTD) simulations quantitatively reproduce the experimental observations with high accuracy revealing the critical roles of layer thicknesses, twist-angle, Fabry–Pérot interference and emitter distribution effects in the determination of polarization conversion efficiency, which yields up to 95 % degrees of circular polarization (DoCP) for the optimized parametrization. Our findings establish a practical route towards fully integrated all-vdW polarization optical components, providing a foundation for nanoscale photonic architectures based entirely on layered materials.**

## Introduction

Since the isolation of graphene[1], great developments in the field of two-dimensional (2D) crystals paved the way to design and control of their electronic, opto-electronic, and magnetic properties through atomically precise fine-tuning of twist-angle, 2D layer thicknesses, composition and arrangement within the assembled few-layer vdW heterostructures[2-7]. Succeeding these developments, transferring nanotechnological recipes applied to graphene towards optically thick multilayers (vdW crystals) of various classes of 2D crystals established a powerful platform for an advancement of visible to infrared (Vis-to-IR) spectral region nanophotonic and integrated optical applications, mostly due to the emerging strong light-matter interaction effects, highly tunable optical responses accompanied by the possession of high refractive indices[8-15].

More recently, the scope of the research in latter directions extended to include multilayer biaxial vdW crystals, further commencing strong polarization-based phenomena[16-18]. This class of vdW crystals exhibits an in-plane optical anisotropy, which arises along with an out-of-plane one (based on a sheer layered structure) from the reduced crystal symmetry. This provides an additional degree of freedom to control and tune the fundamental properties of electromagnetic waves interacting with matter, such as polarization, propagation direction, and phase, making them an attractive platform for realization of ultrathin Vis-*to*-IR polarizing optical components, such as linear polarizers, sub-diffraction half- and quarter- wave plates, or compact waveguides[18-21]. Furthermore, another extension of state-of-the-art recipes applied to graphene towards vdW heterostructures composed of such multilayers, where the same degrees of freedom are intact (tunability of layer thicknesses, twist-angle, composition and arrangement of optically thick layers) further allowed designing more sophisticated polarization based nanoscale optical components and nanophotonic devices, such as beam splitters, chiral circular polarizers, and magic-angle or moiré photonic devices[22-29].

Black phosphorus is an example of a biaxial vdW crystal, which exhibits large in-plane optical anisotropy within nanometer-scale thicknesses due to its direction-dependent dielectric response[30]. Its puckered crystal structure demonstrates an emergence of pronounced linear dichroism between armchair and zigzag axes, paving the way to the control of polarization state[31]. However, even in the few-layer limit, it is not optically transparent. In this sense, research revealing strong in-plane optical anisotropy in low-symmetry multilayers of transition metal or metalloid chalcogenides[18, 22, 32-36], such as GeS, $GeS_2$, $TiS_3$, $AS_2S_3$ made another leap towards the realization of ultrathin polarizing optical components owing to their transparency in the Vis-*to*-IR spectral region. Other examples of vdW crystals exhibiting strong in-plane optical anisotropy that attracted significant attention are the wide-bandgap semiconductors[37, 38], such as $\alpha$-$MoO_3$ and distorted lattice vdW crystals[23, 39-43], such as $ReS_2$ and $ReSe_2$ or oxyhalide family representatives, such as $MoOCl_2$ and $NbOCl_2$. In these systems, an enormous in-plane anisotropy, though typically accompanied by significant optical losses, arises due to the directional Peierls distortion in their crystal lattices, enables sub-diffraction ultraviolet and deep-ultraviolet polarizing optical components.

Despite such rapid progress, recent works mainly focus on free-space optical configurations of those polarizing components, whether in practice, actual nanophotonic applications require miniaturization and direct integration into on-chip optical circuits. In this work, we present a step-*by*-step approach towards the creation of fully integrated all-vdW on-chip polarizing optical components, exemplified by quarter-wave plates operating at near-IR spectral region. Our work establishes a straightforward recipe for an integration of specifically twisted vdW $ReSe_2$/$\alpha$-$MoO_3$ heterostructures on Si/$SiO_2$ substrates, where the in-plane anisotropy of biaxial $\alpha$-$MoO_3$ that serves as quarter-wave plate layer enables a transformation of linearly polarized excitons in $ReSe_2$ into sources of circularly polarized light. Moreover, our FDTD simulations quantitatively reproduce the experimental observations with excellent agreement, elucidating the critical influence of layer thickness, twist-angle, Fabry–Pérot interference, and emitter distribution on the polarization conversion efficiency, where DoCP reaches values as high as 95 % through parameter optimization.

## Results and Discussion

The schematic of the device structure is presented in Figure 1A, where the bottom emitter layer acts as a source of linearly polarized light, while the top biaxially anisotropic layer acts as a quarter-wave plate, in a twisted vdW stack assembled on top of $Si/SiO_2$ substrate. Here, for the emitting source, we chose a multilayer of $ReSe_2$ due to its low-symmetry anisotropic crystal structure (see schematics in Figure 1B, top), forcing excitonic optical dipoles[44,45] to align along a specific in-plane direction, resulting in nearly 100 % of degree of linear polarization (DoLP). An example of a low-temperature photoluminescence (PL) spectrum (see Methods) acquired from bare multilayer of $ReSe_2$ along with its polarization dependence is shown in Figure 1B, bottom. The plot reveals two spectral resonances at 878 nm and 893 nm NIR wavelengths, denoted as $X_1$ and $X_2$, and corresponding to $A$ and $B$ excitonic state emissions with distinct polarization directions exhibiting about 60° rotation mismatch with respect to each other. Next, as a quarter-wave plate layer, we select a multilayer of $\alpha$-$MoO_3$: an in-plane anisotropic vdW crystal transparent in a broad Vis-*to*-IR spectral region, whose optical response is governed by the pronounced structural anisotropy, *i.e.*, orthorhombic crystal structure giving rise to distinct dielectric responses along the three crystallographic directions. The schematics of the crystal structure and the measured optical dispersion of a bare multilayer of $\alpha$-$MoO_3$ are shown in Figure 1C. Its optical constants, extracted from Mueller Matrix (MM) studies (see Methods), reveal pronounced anisotropy across the measured spectral region along with substantial birefringence in both in-plane and out-of-plane directions, in good agreement with previously reported values[46]. In particular, at NIR spectral region of interest (870-920 nm, corresponding to the excitonic emission wavelengths of vdW $ReSe_2$), vdW $\alpha$-$MoO_3$ exhibits an in-plane anisotropy of $\Delta n_{xy} = n_x - n_y \approx 0.08$, which was exploited as a cornerstone to convert linearly polarized excitonic emission of vdW $ReSe_2$ into a circular polarized state.

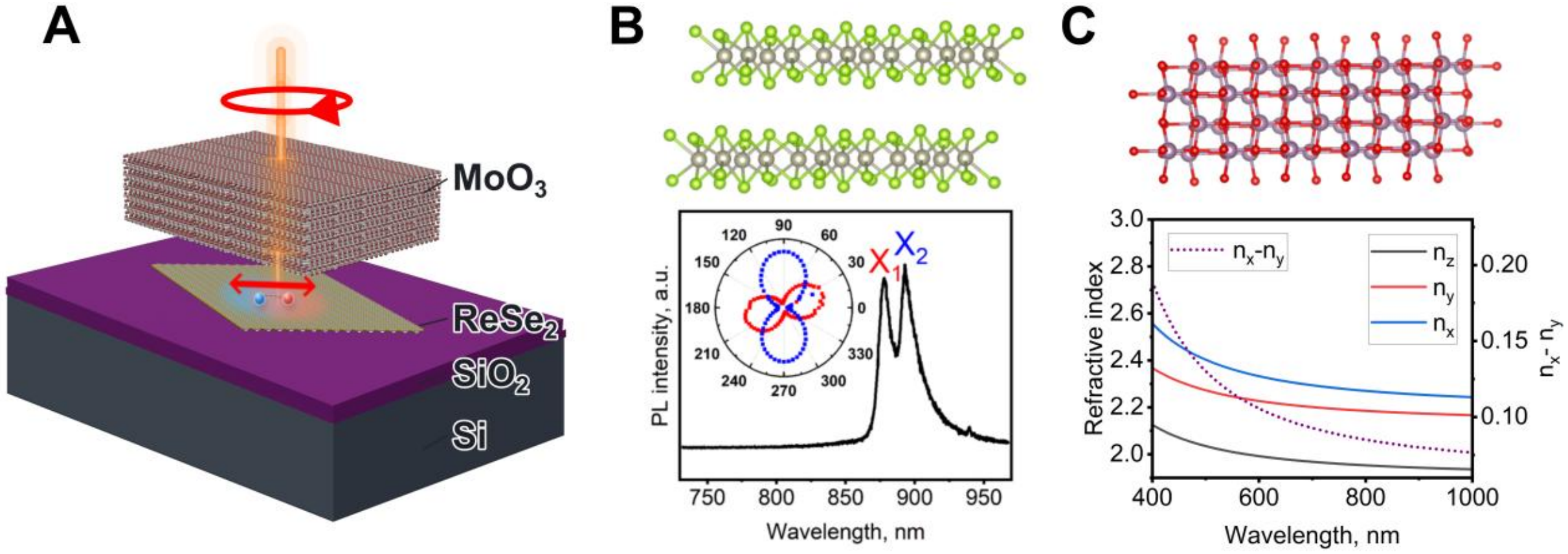

**Figure 1. All-vdW NIR spectral region on-chip quarter-wave plates.** (A) The schematic illustration showing the design of the device composed of pre-aligned $ReSe_2$/$\alpha$-$MoO_3$ vdW heterostructure positioned on top of $Si/SiO_2$ substrate. (B) The schematic illustration of triclinic crystal structure (top, side view) and low-temperature PL spectra (bottom) of vdW $ReSe_2$. The inset represents the angular linear polarization dependence of excitonic states indicated as $X_1$ and $X_2$. (C) The schematic illustration

of orthorhombic atomic structure (top, top view) and experimentally obtained optical constants (bottom) of vdW $\alpha$-$MoO_3$ in 400-1000 nm spectral region.

The control of the alignment between the emitted signal's polarization and the "fast" axis of our vdW $\alpha$-$MoO_3$ quarter-wave plate is crucial for the achievement of a circularly polarized state and enablement of all-vdW on-chip polarization optical components. Therefore, the "fast" axis of vdW $\alpha$-$MoO_3$, *i.e.*, the corresponding in-plane crystallographic direction, should be identified in-prior to the assembly of the vdW stack or controlled *in-situ* in more sophisticated onsite twistable architectures[47,48]. Here, one may employ polarization-resolved Raman spectroscopy (see Methods), whose exemplified spectrum is shown in Figure 2A. Similar to[49], using the strongest $A_g^a$ mode's polarization dependence of vdW $\alpha$-$MoO_3$ in co-polarized configuration, one may describe the angular dependence of its intensity as:

$$R(A_g)\& = \begin{pmatrix} a & 0 & 0 \\ 0 & b & 0 \\ 0 & 0 & c \end{pmatrix}, \quad \boldsymbol{e} = \begin{pmatrix} \cos\theta \\ 0 \\ \sin\theta \end{pmatrix}, \tag{1}$$

$$I_{\parallel}(\theta)\& \propto |\boldsymbol{e}^T R \boldsymbol{e}|^2 \sim \cos^2\theta.$$

The latter exhibits a two-fold symmetry that can be fitted by a simple relation of $cos^2\theta$ with maxima oriented along the *x*-axis as shown in the inset of Figure 2A. Another way to identify the crystallographic axis directions in vdW $\alpha$-$MoO_3$ is through polarization-resolved Vis micro-reflection measurements (see Methods). Here, in the case of a cross-polarized configuration, the intensity of the reflected light can be given as:

$$I_{\perp}(\theta) \propto |A_{\perp}|^2 = \frac{|r_y - r_x|^2}{4} \sin^2 2\theta. \tag{2}$$

where $r_x$ and $r_y$ are reflection coefficients of polarized light along $x$ and $y$ crystallographic axes. Due to an in-plane optical anisotropy of vdW $\alpha$-$MoO_3$ (Figure 1C, bottom), $r_y - r_x \neq 0$, therefore, 4 lobes in the angular dependence are expected with the minima pointing along crystallographic directions (see Figure 2B). Here, the top panel shows an example of reflection micrographs measured at 0°, 20° and 50° with respect to axes directions, demonstrating nearly zero intensity in the case of the former, while revealing a bright image in the case of the latter. The full angular dependence of the reflection intensity along with its fitting for a representative vdW $\alpha$-$MoO_3$ sample are shown in the bottom panel of Figure 2B.

The next critically important parameter in the design of on-chip all-vdW polarizing optical elements is the thickness of our quarter-wave plate, which governs the phase retardance of $\pi/2$ in the two orthogonal directions. This condition is fulfilled when $\frac{2\pi}{\lambda}\ \Delta n_{xy}\ d + \delta_{FP} = \pi/2$, where $d$ is the thickness of an in-plane anisotropic $\alpha$-$MoO_3$ layer, $\Delta\ n_{xy} = n_x - n_y$ is its in-plane birefringence, *i.e.*, the difference between the in-plane components of the refractive index between the two crystallographic axes, and $\delta_{FP}$ is the thickness and proximity media dependent Fabry-Pérot interference contribution to the phase retardance. Here, to elucidate the significant impact of multiple reflections arising in actual on-chip twisted $ReSe_2$/$\alpha$-$MoO_3$ sample (see the design in Figure 1A) on phase retardance, we firstly assume that NIR spectral region electromagnetic wave propagates freely through our in-plane anisotropic layer and evaluate the corresponding phase retardation for air/$\alpha$-$MoO_3$/air structure using recent implementation of Berreman's anisotropic transfer matrix (TMM) method[50, 51]. Supplementary Figure S1 presents the corresponding results with two thicknesses of 2.85 µm and 2.90 µm achieving 90° phase-shift for free standing $\alpha$-$MoO_3$ based quarter-wave plates. Next, we choose representative sub-micron thick vdW $\alpha$-$MoO_3$ quarter-wave plate and $ReSe_2$ emissive multilayer samples for the assembly of our on-chip vdW stack and perform atomic-force microscopy (AFM) scanning (see Methods), whose results are presented in Figure 2C and D. As can be seen, the thicknesses of our vdW samples are of 910 nm for $\alpha$-$MoO_3$, when for $ReSe_2$ we obtain a terraced surface with plateaus of 36, 50, 70 nm in thickness.

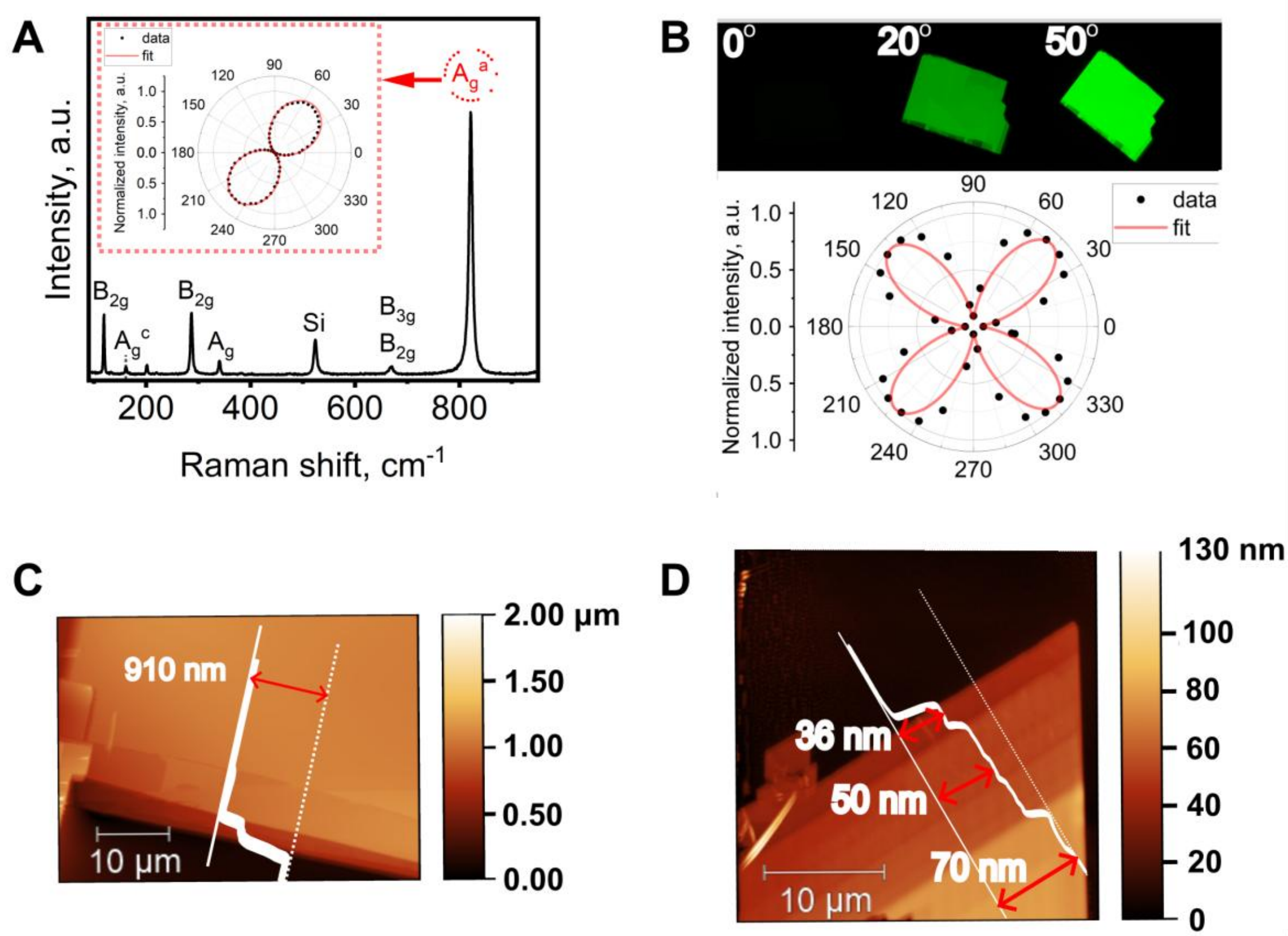

**Figure 2. Identifying crystallographic axes of biaxial α-$MoO_3$ along with thickness characteristics of vdW layers selected for the design of on-chip quarter-wave plates.** (A) The unpolarized Raman spectra of vdW α-$MoO_3$ placed on top of Si/$SiO_2$ substrate illustrating its representative $A_g$ and $B_g$ vibrational modes. The polar diagram in the inset shows the angular polarization dependence of the strongest $A_g^a$ mode obtained in co-polarized geometry. (B) Polarization resolved Vis micro-reflection response of vdW α-$MoO_3$ placed onto Si/$SiO_2$ The top panel displays reflection micrographs at 0°, 20° and 50° rotations with respect to crystallographic axes. The bottom panel shows the full angular dependence of the reflection intensity exhibiting a 4-lobe pattern. (C) and (D) AFM maps along with the thickness profiles (shown by red arrows) corresponding to cross-sections of vdW α-$MoO_3$ and $ReSe_2$ samples initially placed onto Si/$SiO_2$ substrates.

Figure 3 demonstrates the optical micrograph of our proof-of-principle sample, *i.e.*, twisted vdW heterostructure assembled (see Methods) on top of Si/$SiO_2$ substrate for a case of pre-identified directions of linear excitonic polarization in the selected $ReSe_2$ (by measuring polarization-resolved PL response, similar to data shown in the inset of Figure 1B, bottom) and the "fast" axis of the chosen $\alpha$-$MoO_3$ (by measuring polarization-resolved Raman response, similar to data shown in the inset of Figure 2A) with a twist-angle of 21° between them. The polarization-resolved PL mapping (see Figure 3B and Methods) of the resulting twisted vdW stack as a contour plot of DoCP shows that the bare $ReSe_2$ layer demonstrates nearly zero DoCP irrespective of plateau regions of different thicknesses. An example of PL spectra for $\sigma_+$ and $\sigma_-$ circular polarization states acquired from this region is shown in Figure 3C. On the other hand, in our twisted $ReSe_2/\alpha$-$MoO_3$ stack, we observe PL response of excitons to be elliptically polarized. An example of PL spectra from this region for the circular polarization states is presented in Figure 3D. Note the stronger polarizability of $A$ excitonic state for whose wavelength's vicinity we further collect and present the DoCP acquired from the twisted vdW heterostructure region with the thickness of $ReSe_2$ of 70 nm (see Figure 2D) and showing mean elliptical DoCP of 36.6 % (see Figure 3E).

To accurately describe our on-chip proof-of-principle sample's performance, we then implemented FDTD calculations (see Methods) employing the Meep package[52] for our multilayer structure of $Si/SiO_2/ReSe_2/\alpha$-$MoO_3$/air accounting for the measured critical parameters, multiple reflections in the system and an out-of-plane direction distribution of emitters in $ReSe_2$ emerging due to its bulky character ($d$ = 70 nm), which allowed achieving quite high consistency between the measured and simulated DoCPs of 36.6 % and 36.3 % (see Figure 3E and F). Here, the emission wavelength (see Figure 3E), the thicknesses of $ReSe_2$ and $\alpha$-$MoO_3$ layers (see Figure 2C and D), the exact twist-angle between $A$ exciton's linear polarization direction and "fast" axis of $\alpha$-$MoO_3$ (see Figure 3A) along with its in-plane anisotropic optical constants (see Figure 1E) were experimentally defined, leaving only the optical constants of the emissive $ReSe_2$ layer approximated. The latter was applied due to a significant spread in values reported in literature[41, 53-54] accompanied by a recent demonstration of an emergence of Hermitian and non-Hermitian parts in its dielectric tensor[41, 55], which complicates the creation of an exact optical model for this material. Hence, we used, one could say, somewhat better defined, but still approximated optical constants of vdW $ReS_2$, expecting them to lie in the vicinity to that of vdW $ReSe_2$, similar to those based on other S and Se based transition metal dichalcogenides[13]. For the out-of-plane distribution of emitters in $ReSe_2$, we chose exponential shape of their distribution as those are excited optically, with several emitters decreasing in depth $w \propto \exp(-\frac{z}{z_0})$, where $z_0$ was found from a fit: 27 nm, coherent with attenuation coefficient $\frac{4\pi k}{\lambda}$ defined by the extinction coefficient $k$ = 2.6 of the material at the wavelength of interest. The full list of simulation parameters is presented in Supplementary Table S1. Supplementary Figure S2 further shows that at a given thickness of vdW $\alpha$-$MoO_3$ in our on-chip proof-of-principle sample, it would be possible to maximize the DoCP by tuning the twist-angle between $ReSe_2$'s $A$ excitonic state's linear polarization and "fast" axis of $\alpha$-$MoO_3$ achieving a maximum value of 45 %, pointing out the necessity to use thicker $\alpha$-$MoO_3$ layers in real-life applications, which can be potentially attributed to the use of approximated optical constants of $ReS_2$ instead of an elusive optical constants $ReSe_2$. We would also like to point out that such a potential substitution may give rise to a mixture among the fitting parameters, including the out-of-plane distribution of emitters. In the case of the above-mentioned fixed parameterization, the higher DoCP conversion yield exceeding 95 % can be achieved for quarter-wave plate thicknesses above 1.3 µm (see Supplementary Figure S2 for details).

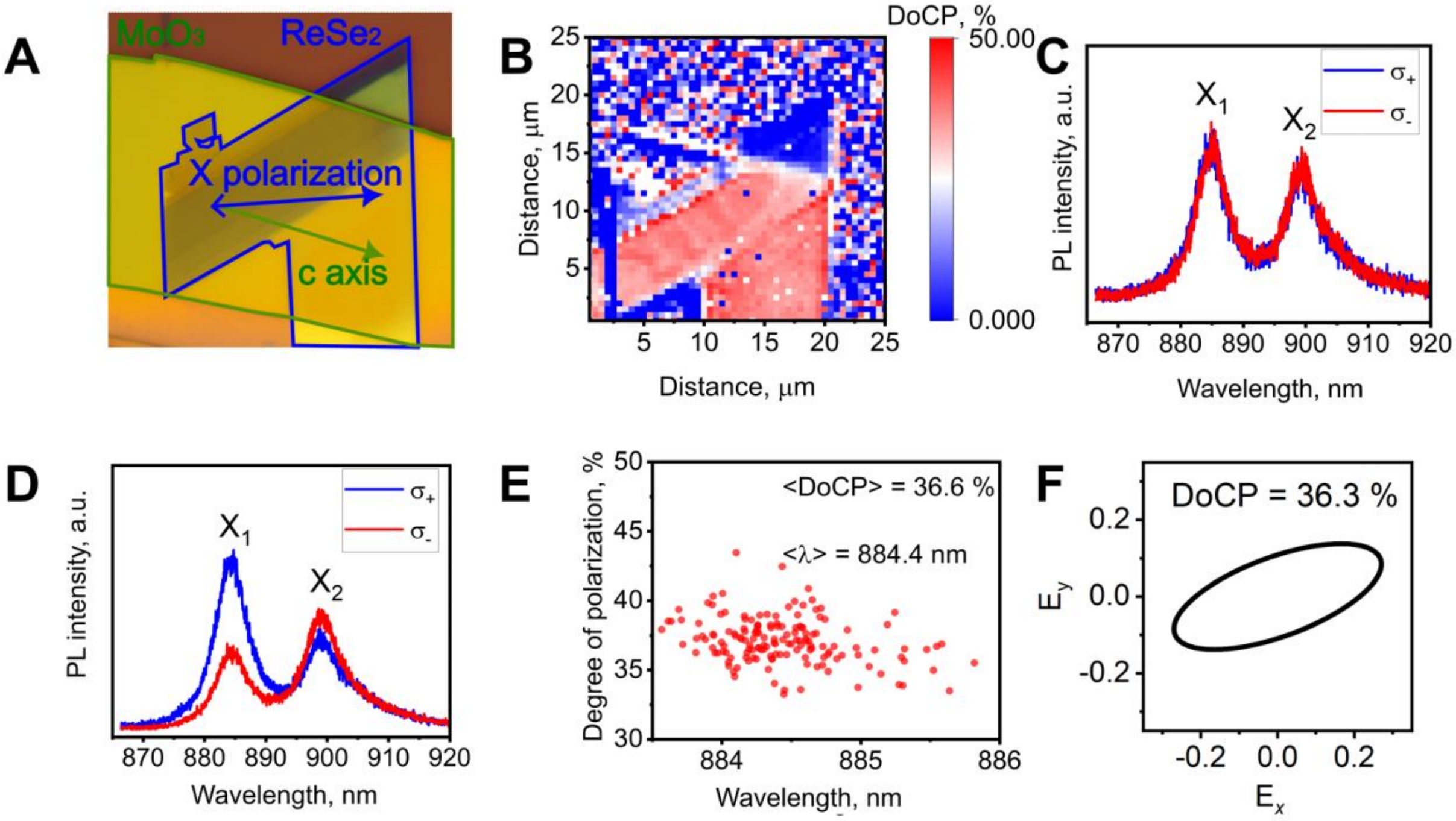

**Figure 3. Probing DoCP in $ReSe_2$/α-$MoO_3$twisted stack assembled on top of Si/$SiO_2$ chip.** (A) 50 X optical micrograph of our proof-of-principle sample. The blue arrow denotes the linear polarization direction of $A$ excitonic emission from $ReSe_2$, while the green arrow shows the "fast" axis of α-$MoO_3$. (B) The DoCP map collected all over the sample, including bare $ReSe_2$ terraces and substrate. (C) and (D) Examples of circularly polarized $\sigma_+$ and $\sigma_-$ state PL spectra acquired from bare $ReSe_2$ sample of 70 nm thickness (C) and $ReSe_2$/α-$MoO_3$ twisted heterostructure (D). The excitonic states are denoted as $X_1$ and $X_2$. (E) The DoCP histogram collected from twisted stack as a function of wavelength (in a close vicinity to $A$ excitonic emission energy). (F) FDTD evaluated DoCP showing the trajectory of the electric field's vector in $E_{xy}$ plane.

## Conclusion

To sum up, we presented a direct route towards the creation of on-chip polarizing optical components based on all-vdW twisted heterostructures composed of emissive $ReSe_2$ layer, which served as a source of NIR spectral region linearly polarized excitonic emission, and a multilayer of $\alpha$-$MoO_3$, which was used as quarter-wave plate layer. Here, DoCP was probed by polarization-resolved PL mapping, which presented a clear contrast between the bare $ReSe_2$ and our proof-of-principle twisted sample's excitonic responses. A consistent FDTD optical model describing arising elliptical polarization was constructed and affirmed by utilization of optical constants of vdW $\alpha$-$MoO_3$ measured through spectroscopic micro-ellipsometry. Our evaluations further showed that from the standpoint of $\alpha$-$MoO_3$ based quarter-wave plate layer's characteristics (assembled into the twisted vdW stack), to achieve a near-perfect circular polarization state, one would require either operation at lower wavelengths, where in-plane anisotropy of $\alpha$-$MoO_3$ is larger, or a thickening of the layer over 1.3 μm. We believe that our findings will be of great use for the understanding of peculiarities in the design of on-chip polarizing optical components based on all-vdW twisted heterostructures, and can be potentially extended not only to other biaxial and low-symmetry vdW crystals, but also applied for a conversion of linearly polarized single photonic emission arising at the surface or bulk of some vdW crystals.

## Methods

**Sample fabrication.** $ReSe_2$ and $\alpha$-$MoO_3$ crystals were purchased from 2D Semiconductors and micro-mechanically cleaved into optically thick multilayer samples on top of standard Si/$SiO_2$ substrates at room temperature using commercial scotch tapes from Nitto Denko Corporation. Prior to micro-mechanical cleavage, the substrates were sequentially cleaned in acetone, isopropyl alcohol and deionized water, followed by an air plasma treatment to remove residual surface contaminants and improve the adhesion. The twisted $ReSe_2$/$\alpha$-$MoO_3$ sample was prepared using a commercial PDMS stamp-based deterministic dry-transfer technique, where the individual layers were aligned with respect to each other by utilization of micro-mechanical control over a twist angle during the stacking.

**Spectroscopic micro-ellipsometry.** To accurately access the dielectric tensor components, biaxial vdW $\alpha$-$MoO_3$ samples were studied on 295 nm Si/$SiO_2$ and bare Si substrates with their thickness preliminary determined by AFM studies. The anisotropic optical response was then measured at room temperature using MM imaging spectroscopic micro-ellipsometry with a Park Systems Accurion EP4 imaging ellipsometer over 400-940 nm spectral region at different angles of incidence (AoI). To further support the subsequent fitting procedure and reduce parameter correlations, the experimental dataset was built from vdW $\alpha$-$MoO_3$ samples of different thicknesses and arbitrary in-plane orientations on both substrates. Additional MM measurements were collected at 633 nm and AoI of 55$^o$ across 20 azimuthal orientations spanning 0–180$^o$. The measured data were then analyzed in EP4Model software using an anisotropic multilayer optical model, in which vdW $\alpha$-$MoO_3$ was treated as a biaxial material with two in-plane and one out-of-plane dielectric tensor components. Given the negligible Vis spectral region absorption, its optical response was treated as lossless and both the in-plane and out-of-plane components were described using Sellmeier's dispersion relation: $n_i^2(\lambda)$ = 1 + $B_i\ \lambda^2/(\lambda^2 - C_i)$, where $i$ = $x, y, z$ (Sellmeier's coefficients obtained from the fit are summarized in Supplementary Table S2).

**Polarization-resolved Raman spectroscopy.** Raman spectra from vdW $\alpha$-$MoO_3$ samples were acquired using a custom designed system comprising a confocal microscope equipped with a 100 X objective lens (N.A. = 0.70) and PyLoN 100-BRX LN2 cooled CCD camera coupled to a SpectraPro HRS-750 spectrometer. The measurements were carried out using a 532 nm CW laser and RazorEdge LP03-532RE-25 long pass filter. The polarization angle was varied continuously from 0$^o$ to 360$^o$ with a step of 5 $^o$ by simultaneous rotation of half-wave plates in the excitation and detection paths, while in the detection path, the position of the polarizer was fixed.

**Micro-reflection spectroscopy.** Vis micro-reflection spectroscopy was performed by utilizing a commercial optical microscope (Nikon Eclipse LV100ND) equipped with polarizers in the excitation and detection paths aligned at 90$^o$ and a green bandpass filter in the excitation path. To control the rotation angle of the investigated samples, a rotational mount from Thorlabs components was used.

**Polarization-resolved PL spectroscopy.** To perform polarization-resolved PL measurements, the samples were loaded into a dry cryogenic system with a base temperature of 4 K and mounted onto *x-y-z* piezo-stages. The measurements were realized in a back-scattering geometry, and the laser focusing and collection were done by a cold objective with N.A. = 0.82. For the detection of the signal, the light was dispersed by a SpectraPro HRS-750 spectrometer (Princeton Instruments) and detected by a PyLoN 100-BRX LN2 cooled CCD camera (Princeton Instruments). Depolarized light was used for the measurements of DoCP signals. For the collection of the signal, a quarter-wave plate was placed in the sequence with a polarizer in front of the spectrometer. A half-wave plate was placed in the sequence with a polarizer in front of the spectrometer for the collection of DoLP signals.

**Atomic-force microscopy.** Topography measurements were performed using Bruker's Dimension FastScan XR atomic-force microscope operated in tapping and peak force modes with RTESPA-150 cantilevers.

**TMM and FDTD calculations.** TMM studies were employed to estimate the thickness of bare $\alpha$-$MoO_3$ layers used as quarter-wave plates for a system containing 3 layers: air/$\alpha$-$MoO_3$/air, where the optical constants of $\alpha$-$MoO_3$ were taken according to the ellipsometry fit at a wavelength of 884 nm: $n_x$ = 2.26, $n_y$ = 2.17, $n_z$ = 1.94. FDTD calculations were performed for experimental parameters describing our proof-of-principle multilayer Si/$SiO_2$/$ReSe_2$/$\alpha$-$MoO_3$/air sample. Here, the optical model consisted of 295 nm thick $SiO_2$, 70 nm thick $ReSe_2$ and 910 nm thick $\alpha$-$MoO_3$ layers placed onto Si substrate. The optical constants of latter were taken in accordance with an ellipsometric fit using the same parameters as in TMM calculations. The twist-angle between the linearly polarized excitonic emission and the "fast" axis of vdW $\alpha$-$MoO_3$ was taken as 21°. An exponentially decaying amplitude envelope for the light source with the decay length of 27 nm was introduced to describe the out-of-plane distribution of emitters. The space-grid had 1 nm resolution, and the time step was 6.6 $10^{-19}$ s. The field monitor was placed at 3 µm above the vdW $\alpha$-$MoO_3$ layer, capturing $E_x$ and $E_y$ components every 0.07 fs, which resulted in about 50 points per one period at a given frequency. The resulting trajectory of the electric field's vector in the $E_{xy}$ plane after multiple reflections had the shape of an ellipse, which defined the DoCP as the ratio between its main axes.

## References

[1] Novoselov Kostya S, Andre K Geim, Sergei V Morozov, et al. Electric Field Effect in Atomically Thin Carbon Films. Science 306 (5696): 666–69 (2004).

[2] Geim Andre K, and Irina V Grigorieva. Van Der Waals Heterostructures. Nature 499 (7459): 419–25 (2013).

[3] Cao Yuan, Valla Fatemi, Shiang Fang, et al. Unconventional Superconductivity in Magic-Angle Graphene Superlattices. Nature 556 (7699): 43–50 (2018).

[4] Mak Kin Fai, Changgu Lee, James Hone, Jie Shan, and Tony F Heinz. Atomically Thin MoS2: A New Direct-Gap Semiconductor." Physical Review Letters 105 (13): 136805 (2010).

[5] Huang Bevin, Genevieve Clark, Efrén Navarro-Moratalla, et al. Layer-Dependent Ferromagnetism in a van Der Waals Crystal down to the Monolayer Limit." Nature 546 (7657): 270–73 (2017).

[6] Tran Kha, Galan Moody, Fengcheng Wu, et al. Evidence for Moiré Excitons in van Der Waals Heterostructures." Nature 567 (7746): 71–75 (2019).

[7] Wang Qing Hua, Kourosh Kalantar-Zadeh, Andras Kis, Jonathan N Coleman, and Michael S Strano. Electronics and Optoelectronics of Two-Dimensional Transition Metal Dichalcogenides. Nature Nanotechnology 7 (11): 699–712 (2012).

[8] Low Tony, Andrey Chaves, Joshua D Caldwell, et al. Polaritons in Layered Two-Dimensional Materials. *Nature Materials* 16 (2): 182–94 (2017).

[9] Basov D N, M M Fogler, and F J García de Abajo. Polaritons in van Der Waals Materials. *Science* 354 (6309): aag1992 (2016).

[10] Koppens Frank H L, Thomas Mueller, Ph Avouris, Andrea C Ferrari, M Serena Vitiello, and Marco Polini. Photodetectors Based on Graphene, Other Two-Dimensional Materials and Hybrid Systems. *Nature Nanotechnology* 9 (10): 780–93 (2014).

[11] Britnell Liam, Ricardo Mendes Ribeiro, Axel Eckmann, et al. Strong Light-Matter Interactions in Heterostructures of Atomically Thin Films. *Science* 340 (6138): 1311–14 (2013).

[12] Mak Kin Fai, and Jie Shan. Photonics and Optoelectronics of 2D Semiconductor Transition Metal Dichalcogenides. *Nature Photonics* 10 (4): 216–26 (2016).

[13] Munkhbat Battulga, Piotr Wróbel, Tomasz J. Antosiewicz, and Timur O. Shegai. Optical Constants of Several Multilayer Transition Metal Dichalcogenides Measured by Spectroscopic Ellipsometry in the 300–1700 nm Range: High Index, Anisotropy, and Hyperbolicity. *ACS Photonics* 9 (7): 2398–407 (2022).

[14] Abajo F. Javier García de, D. N. Basov, Frank H. L. Koppens, et al. Roadmap for Photonics with 2D Materials. *ACS Photonics* 12 (8): 3961–4095 (2025).

[15] Ren Wencai, Peter Bøggild, Joan Redwing, et al. The 2D Materials Roadmap. *2D Materials* 13 (2): 021501 (2026).

[16] Ma Weiliang, Pablo Alonso-González, Shaojuan Li, et al. In-Plane Anisotropic and Ultra-Low-Loss Polaritons in a Natural van Der Waals Crystal. *Nature* 562 (7728): 557–62 (2018).

[17] Zheng Zebo, Ningsheng Xu, Stefano L. Oscurato, et al. A Mid-Infrared Biaxial Hyperbolic van Der Waals Crystal. *Science Advances* 5 (5): eaav8690 (2019).

[18] Slavich Aleksandr S., Georgy A. Ermolaev, Mikhail K. Tatmyshevskiy, et al. Exploring van Der Waals Materials with High Anisotropy: Geometrical and Optical Approaches. *Light: Science & Applications* 13: 1 (2024).

[19] Abedini Dereshgi, Sina, Yea-Shine Lee, Maria Cristina Larciprete, Marco Centini, Vinayak P Dravid, and Koray Aydin. Low-Symmetry $\alpha$-$MoO_3$ Heterostructures for Wave Plate Applications in Visible Frequencies. *Advanced Optical Materials* 11 (7): 2202603 (2023).

[20] Yang He, Henri Jussila, Anton Autere, et al. Optical Waveplates Based on Birefringence of Anisotropic Two-Dimensional Layered Materials. *ACS Photonics* 4 (12): 3023–30 (2017).

[21] Li Zhipeng, Xuezhi Ma, Fengxia Wei, et al. As-Grown Miniaturized True Zero-Order Waveplates Based on Low-Dimensional Ferrocene Crystals. *Advanced Materials* 35 (32): 2302468 (2023).

[22] Slavich Aleksandr S., Georgy A. Ermolaev, Ilya A. Zavidovskiy, et al. Germanium Disulfide as an Alternative High Refractive Index and Transparent Material for UV-Visible Nanophotonics. *Light: Science & Applications* 14: 213 (2025).

[23] Margaryan A. V., M. L. Sargsyan, M. H. Hayrapetyan, D. A. Karakhanyan, K. S. Novoselov, and D. A. Ghazaryan. Dynamic Dielectric Permittivity Tensor of in-Plane Hyperbolic van Der Waals $MoOCl_2$ and Emergent Chiral Photonic Applications. *Npj 2D Materials and Applications 10*: 42 (2026).

[24] Huang Shuxin, Guanyu Zhang, Yaolong Li, et al. Ultrathin, High-Performance Circular Wave Plate Based on Natural van Der Waals Crystals. *Research Square* (2026).

[25] Enders Michael T., Mitradeep Sarkar, Evgenia Klironomou, et al. Mid-Infrared Chirality and Chiral Thermal Emission from Twisted $\alpha$-$MoO_3$. *Nature Communications* 16: 1 (2025).

[26] Hu Guangwei, Qingdong Ou, Guangyuan Si, et al. Topological Polaritons and Photonic Magic Angles in Twisted $\alpha$-$MoO_3$ Bilayers. *Nature* 582 (7811): 209–13 (2020).

[27] Tang Haoning, Xueqi Ni, Fan Du, Vishantak Srikrishna, and Eric Mazur. On-Chip Light Trapping in Bilayer Moiré Photonic Crystal Slabs. *Applied Physics Letters* 121 (23): 231702 (2022).

[28] Du Luojun, Maciej R. Molas, Zhiheng Huang, et al. Moiré Photonics and Optoelectronics. *Science* 379 (6639): eadg0014 (2023).

[29] Zhang Mingjie, and Eric Mazur. Dielectric Photonic Moiré Metamaterials. *Npj Metamaterials* 2: 20 (2026).

[30] Xia Fengnian, Han Wang, and Yichen Jia. Rediscovering Black Phosphorus as an Anisotropic Layered Material for Optoelectronics and Electronics. *Nature Communications* 5, 1: 4458 (2014).

[31] Wang Xiaomu, Aaron M Jones, Kyle L Seyler, et al. Highly Anisotropic and Robust Excitons in Monolayer Black Phosphorus. *Nature Nanotechnology* 10 (6): 517–21 (2015).

[32] Chenet Daniel A, Burak Aslan, Pinshane Y Huang, et al. In-Plane Anisotropy in Mono-and Few-Layer $ReS_2$ Probed by Raman Spectroscopy and Scanning Transmission Electron Microscopy. *Nano Letters* 15 (9): 5667–72 (2015).

[33] Arora Ashish, Jonathan Noky, Matthias Drüppel, et al. Highly Anisotropic in-Plane Excitons in Atomically Thin and Bulklike 1T'-$ReSe_2$. *Nano Letters* 17 (5): 3202–7 (2017).

[34] Li Zongbao, Yusi Yang, Xia Wang, Wei Shi, Ding-Jiang Xue, and Jin-Song Hu. Three-Dimensional Optical Anisotropy of Low-Symmetry Layered GeS. *ACS Applied Materials & Interfaces* 11 (27): 24247–53 (2019).

[35] Liu Huili, Xiaoxia Yu, Kedi Wu, et al. Extreme in-Plane Thermal Conductivity Anisotropy in Titanium Trisulfide Caused by Heat-Carrying Optical Phonons. *Nano Letters* 20 (7): 4891–98 (2020).

[36] Zhang Jiwei, Wenrui Li, Jiahao Li, et al. Quantitative Determination of in-Plane Optical Anisotropy by Surface Plasmon Resonance Holographic Microscopy. *Light: Science & Applications* 15 (1): 152 (2026).

[37] Puebla Sergio, Roberto D'Agosta, Gabriel Sanchez-Santolino, Riccardo Frisenda, Carmen Munuera, and Andres Castellanos-Gomez. In-Plane Anisotropic Optical and Mechanical Properties of Two-Dimensional $MoO_3$. *Npj 2D Materials and Applications* 5 (1): 37 (2021).

[38] Lajaunie L., F. Boucher, R. Dessapt, and P. Moreau. Strong Anisotropic Influence of Local-Field Effects on the Dielectric Response of $\alpha$-$MoO_3$. *Physical Review B* 88 (11): 115141 (2013).

[39] Gao Jia, Caokun Wang, Chorng Haur Sow, Qi Zhang, and Eng Tuan Poh. Ultrathin Quarter-Waveplates Based on Two-Dimensional Anisotropic $NbOCl_2$. *Nature Communications* 17: 4118 (2026).

[40] Ermolaev Georgy, Adilet Toksumakov, Valeria Maslova, et al. Deep-Subwavelength and Broadband Quarter-Wave Retardation in Ultrathin Hyperbolic $MoOCl_2$. *arXiv Preprint arXiv:2604.05236* (2026).

[41] Ermolaev Georgy A., Kirill V. Voronin, Adilet N. Toksumakov, et al. Wandering Principal Optical Axes in van Der Waals Triclinic Materials. *Nature Communications* 15 (1): 1552 (2024).

[42] Guo Qiangbing, Qiuhong Zhang, Tan Zhang, et al. Colossal in-Plane Optical Anisotropy in a Two-Dimensional van Der Waals Crystal. *Nature Photonics* 18: 1170–75 (2024).

[43] Ermolaev Georgy, Adilet Toksumakov, Aleksandr Slavich, et al. Giant Optical Anisotropy and Visible-Frequency Epsilon-Near-Zero in Hyperbolic van Der Waals $MoOCl_2$. *Nano Letters* 26 (13): 4329–38 (2026).

[44] Aslan Ozgur Burak, Daniel A. Chenet, Arend M. van der Zande, James C. Hone, and Tony F. Heinz. Linearly Polarized Excitons in Single- and Few-Layer $ReS_2$ Crystals. *ACS Photonics* 3 (1): 96–101 (2016).

[45] Ahn Jongtae, Kyul Ko, Ji-Hoon Kyhm, et al. Near-Infrared Self-Powered Linearly Polarized Photodetection and Digital Incoherent Holography Using $WSe_2$/$ReSe_2$ van Der Waals Heterostructure. *ACS Nano* 15 (11): 17917–25 (2021).

[46] Andres-Penares Daniel, Mauro Brotons-Gisbert, Cristian Bonato, Juan F. Sánchez-Royo, and Brian D. Gerardot. Optical and Dielectric Properties of $MoO_3$ Nanosheets for van Der Waals Heterostructures. *Applied Physics Letters* 119 (22): 223104 (2021).

[47] Ribeiro-Palau Rebeca, Changjian Zhang, Kenji Watanabe, Takashi Taniguchi, James Hone, and Cory R. Dean. Twistable Electronics with Dynamically Rotatable Heterostructures. *Science* 361 (6403): 690–93 (2018).

[48] Yang Yaping, Jidong Li, Jun Yin, et al. In Situ Manipulation of van Der Waals Heterostructures for Twistronics. *Science Advances* 6 (49): eabd3655 (2020).

[49] Gong Yong, Yuxuan Zhao, Zhen Zhou, et al. Polarized Raman Scattering of in-Plane Anisotropic Phonon Modes in $\alpha$-$MoO_3$. *Advanced Optical Materials* 10 (10): 2200038 (2022).

[50] Berreman Dwight W. Optics in Stratified and Anisotropic Media: 44-Matrix Formulation. *J. Opt. Soc. Am.* 62 (4): 502–10 (1972).

[51] Passler Nikolai Christian, and Alexander Paarmann. Generalized 4  4 Matrix Formalism for Light Propagation in Anisotropic Stratified Media: Study of Surface Phonon Polaritons in Polar Dielectric Heterostructures. *J. Opt. Soc. Am. B* 34 (10): 2128–39 (2017).

[52] Oskooi Ardavan F., David Roundy, Mihai Ibanescu, Peter Bermel, J. D. Joannopoulos, and Steven G. Johnson. Meep: A Flexible Free-Software Package for Electromagnetic Simulations by the FDTD Method." *Computer Physics Communications* 181 (3): 687–702 (2010).

[53] Shubnic Anton A., Roman G. Polozkov, Ivan A. Shelykh, and Ivan V. Iorsh. High Refractive Index and Extreme Biaxial Optical Anisotropy of Rhenium Diselenide for Applications in All-Dielectric Nanophotonics. *Nanophotonics* 9 (16): 4737–42 (2020).

[54] Mikhin Alexey, Anton Shubnic, Tatiana Ivanova, Ivan Shelykh, Anton K. Samusev, and Ivan Iorsh. Bulk $ReSe_2$: Record High Refractive Index and Biaxially Anisotropic Material for All-Dielectric Nanophotonics." *ACS Photonics* 10 (6): 1769–74 (2023).

[55] Chakrabarty Devarshi, Avijit Dhara, Pritam Das, Kritika Ghosh, Ayan Roy Chaudhuri, and Sajal Dhara. Anisotropic Exciton-Polaritons Reveal Non-Hermitian Topology in van Der Waals Materials. *Nano Letters* 26 (12): 4089–95 (2026).